%% file: main.tex
\documentclass[conference]{IEEEtran}
\IEEEoverridecommandlockouts
\usepackage{array}
\usepackage{soul}
\usepackage[font=footnotesize]{subcaption}

\usepackage[font=footnotesize]{caption}
\usepackage{comment}
\usepackage{amsmath,amssymb,amsfonts}
\usepackage{algorithmic}
\usepackage{graphicx}
\usepackage{textcomp}
\usepackage{float}
\usepackage{xcolor}
\usepackage{multicol}
\usepackage{multirow}
\def\BibTeX{{\rm B\kern-.05em{\sc i\kern-.025em b}\kern-.08em
    T\kern-.1667em\lower.7ex\hbox{E}\kern-.125emX}}
\begin{document}

\title{High quality ECG dataset based on MIT-BIH recordings for improved heartbeats classification
}

\IEEEoverridecommandlockouts
\IEEEpubid{\makebox[\columnwidth]{979-8-3503-4647-3/23/\$31.00~\copyright2023 IEEE \hfill} \hspace{\columnsep}\makebox[\columnwidth]{ }}

\author{

\IEEEauthorblockN{Ahmed.S Benmessaoud}
\IEEEauthorblockA{
\textit{Innovation Academy Mila}\\
Mila, Algeria \\
ahmed.sif.benmessaoud.13@gmail.com}

\and

\IEEEauthorblockN{Farida Medjani}
\IEEEauthorblockA{\textit{Mathematics and their interactions Laboratory} \\
\textit{Abdelhafid Boussouf University Center}\\
Mila, Algeria \\
 f.medjani@centre-univ-mila.dz}

\and

\IEEEauthorblockN{Yahia Bousseloub}
\IEEEauthorblockA{
\textit{Electromechanical Department}\\
Badji Mokhtar-Annaba University\\
Annaba, Algeria \\
yahia.bousseloub@univ-annaba.org}

\and

\IEEEauthorblockN{Khalid Bouaita}
\IEEEauthorblockA{
\textit{Innovation Academy Mila}\\
Mila, Algeria \\
khatec@gmail.com}

\and

\IEEEauthorblockN{Dhia Benrahem}
\IEEEauthorblockA{
\textit{Innovation Academy Mila}\\
Mila, Algeria \\
dhiaben048@gmail.com}

\and

\IEEEauthorblockN{Tahar Kezai}
\IEEEauthorblockA{
\textit{IEEE member}\\
\textit{Innovation Academy Mila}\\
Mila, Algeria \\
tkezai@gmail.com}

}

\maketitle

\IEEEpubidadjcol

\begin{abstract}
Electrocardiogram (ECG) is a reliable tool for medical professionals to detect and diagnose abnormal heart waves that may cause cardiovascular diseases. This paper proposes a methodology to create a new high-quality heartbeat dataset from all 48 of the MIT-BIH recordings. The proposed approach computes an optimal heartbeat size, by eliminating outliers and calculating the mean value over 10-second windows. This results in independent QRS-centered heartbeats avoiding the mixing of successive heartbeats problem. The quality of the newly constructed dataset has been evaluated and compared with existing datasets. To this end, we built and trained a PyTorch 1-D Resnet architecture model that achieved 99.24\% accuracy with a 5.7\% improvement compared to other methods. Additionally, downsampling the dataset has improved the model's execution time by 33\% and reduced 3x memory usage.

%
\end{abstract}

\begin{IEEEkeywords}
ECG, deep learning, heartbeat, PhysioBank MIT-BIH arrhythmia database,convolutional neural network, resnet
\end{IEEEkeywords}

\input{sections/0intro}
\input{sections/1methods}

\input{sections/2model}
\input{sections/3results}
\input{sections/4conc}


\bibliographystyle{IEEEtran}
\bibliography{library}

\end{document}

%% file: sections/0intro.tex
\section{Introduction}

Heart disease is a major global cause of death, with heart arrhythmia accounting for 16\% of deaths in the last two decades (World Health Organization, 2019) \cite{who}. ECG signals are widely used for diagnosing heart disease due to their affordability, convenience, ease of use, and precision. Support Vector Machine (SVM), Multilayer Perceptron (MLP), and Decision Tree (DT) algorithms have been commonly used in machine learning to classify and detect anomalies in ECG signals \cite{4015610, 5238547, 7491263}. Recent research in machine learning (ML) has found that Convolutional Neural Network (CNN)-based automated feature extraction techniques are scalable and highly accurate in ECG signal classification. Deep learning approaches, including those in \cite{ACHARYA2017389, Kachuee, https://doi.org/10.48550/arxiv.1707.01836}, have achieved state-of-the-art results, even competing with human cardiologists in signal analysis.

However, deep learning requires large amounts of data and computation and the usage of GPUs. Moreover, the performance of deep learning models is highly impacted by the quality of data. Thus, effective preprocessing techniques are crucial to enhance data quality, reduce data dimensionality, and minimize both model size and training and inference time. One of the challenges in ECG heartbeat classification is the limited number of public datasets available. The two largest publicly accessible datasets are the MIT \cite{mitbih} and PTB \cite{PTB} datasets, which provide only continuous ECG recordings on the time domain and their corresponding labels.

\section{Motivations}

 We have found only two references that have proposed methodologies to improve the quality of the MIT recordings by creating new heartbeat datasets (Acharaya \textit{et al.} \cite{ACHARYA2017389} and Kachuee \textit{et al.} \cite{Kachuee}). Unfortunately, only \cite{Kachuee} has made their dataset publicly available. The goal of this work is to create and publish a new high-quality heartbeat dataset based on MIT recordings. We selected the MIT dataset, due to its large size, diversity, and comprehensive annotations, compared to the PTB dataset. 

Recently, Acharya \textit{et al.} \cite{ACHARYA2017389} have segmented all the MIT dataset ECG recordings into segments of 260 samples and centered them around their R-peaks. This approach ensures that important information in each heartbeat is preserved and not mixed with those of surrounding heartbeats. However, the authors' choice of 260 samples as heartbeat size, based on a 360Hz sampling rate, is problematic. First, they have not given a method that permits estimating this value. Secondly, we found that in numerous ECG recordings from the MIT dataset, over 50\% of heartbeats have a length greater than 260 samples, as illustrated in Fig. \ref{fig:box_plot}. This highlights the need for a more comprehensive analysis to estimate an appropriate heartbeat size that must include the maximum heartbeat signal information.

\begin{figure}[htbp]
    \centering
    \includegraphics[width=3cm]{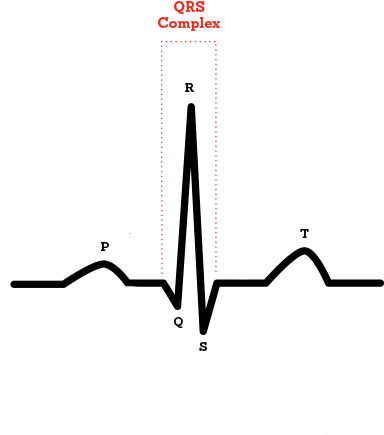}
    \caption{Representation of the P, QRS, T wave of a single heartbeat in an ECG recording}
    \label{fig:heartbeartQRS}
\end{figure}

On the other hand, Kachuee \textit{et al.} \cite{Kachuee} have published their own dataset of heartbeats based on R-R intervals (from the MIT dataset). They extracted each heartbeat starting from the previous R peak to 1.2$T$, where $T$ is the median of the R-R intervals in a 10s window. Indeed, this approach improved the issue of heartbeats containing only 260 samples. Regrettably, adding the RS, T waves of the previous heartbeat to the P, QRS of the current heartbeat (Fig. \ref{fig:heartbeartQRS}) leads to mixing samples of the two consecutive heartbeats. This highly impacts the quality of the dataset, especially in cases where successive heartbeats have different labels.

In this paper, we mainly focused on creating heartbeats that are limited to their own P, QRS, T waves. To overcome the mixing problem, we propose a new methodology to create a high-quality heartbeat dataset based on MIT recordings. We first eliminated outlier heartbeats using the IQR method \cite{IQR} (see section \ref{out}). Then, adaptive heartbeat size has been computed, considering the mean value of RR time intervals over the current 10-second window. This resulted in QRS-centered heartbeats containing only one heartbeat including its own waves, avoiding the problem of mixing successive heartbeats. We also preserved the R-R time interval set, which is a crucial diagnostic parameter for certain diseases, including atrial fibrillation. Subsequently, we developed a 1-D Resnet architecture model to evaluate the performance and quality of the new dataset compared to work done by Acharya \textit{et al.} \cite{ACHARYA2017389} and Kachuee \textit{et al.} \cite{Kachuee}. Our main contributions are the prepossessing methodology to generate a clean dataset ready for classification, and a deep 1-D resnet architecture that achieves state of the art performance. The following sections present the details of the proposed methodology and show the quality of the experiments' results.

%% file: sections/1methods.tex
\section{Dataset}
\label{sec:dataset}

In the late 1970s, the BIH Arrhythmia Laboratory conducted a study on 47 subjects, resulting in 48 half-hour ECG recordings. These recordings, digitized at 360 samples per second per channel, were reviewed and annotated by at least two cardiologists, ensuring the accuracy of the approximately 110,000 annotations. These heartbeats were then categorized into 24 subcategories, which were further grouped into 5 main categories as described in table \ref{table:AnnotationD}

\begin{table}[H]
    \centering
        \caption{Data distribution of each class}

    \begin{tabular}{c c c c c c}
        \hline
        N & Q & V & S & F & \textbf{Total}\\
        \hline
        90,631&8,043 & 7,236 & 2,781 & 803 & \textbf{109,494}\\
        82.77\%&7.35\% & 6.61\%& 2.54\% & 0.73\% & \textbf{100\%}\\
        \hline
    \end{tabular} 
        \label{table:AnnotationD}

\end{table}

\subsection{Outlier removal}
\label{out}

\begin{figure}[H]
    \centering
    \includegraphics[width=6cm]{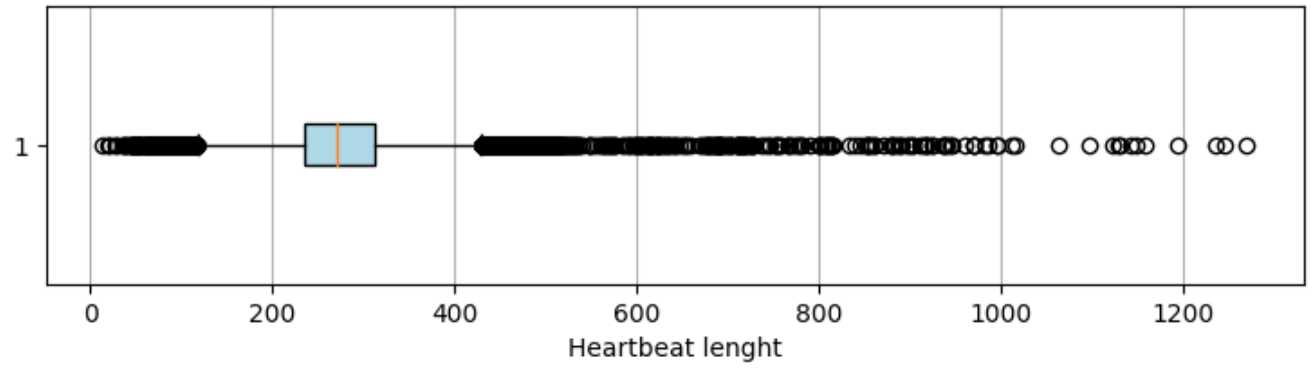}
    \caption{Heartbeat length distribution visualized using a box plot}
    \label{fig:box_plot}
\end{figure}

From the set of RR time intervals lengths a box plot is created to visualize their distribution and identify outliers (Fig. \ref{fig:box_plot}). 

As shown in Fig. \ref{fig:box_plot}, we have defined outliers as heartbeats with sizes that deviate significantly from the majority of the data.

\begin{figure}[H]
  \centering
  \begin{subfigure}{0.21\textwidth}
  \centering
    \includegraphics[width=\textwidth]{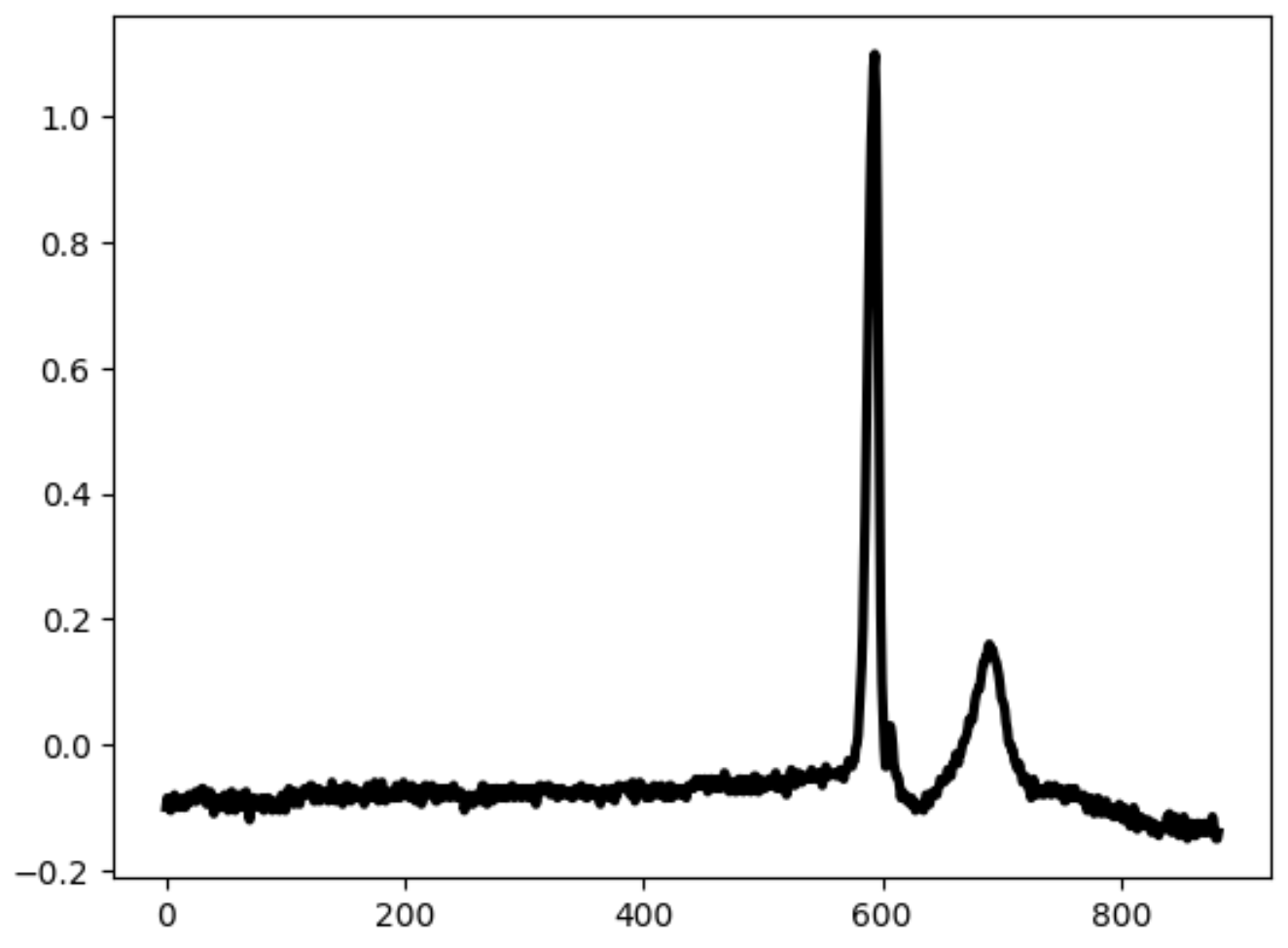}
    \caption{Upper outlier}
    \label{fig:out1}
  \end{subfigure}
  \hfill
  \begin{subfigure}{0.21\textwidth}
    \centering
    \includegraphics[width=\textwidth]{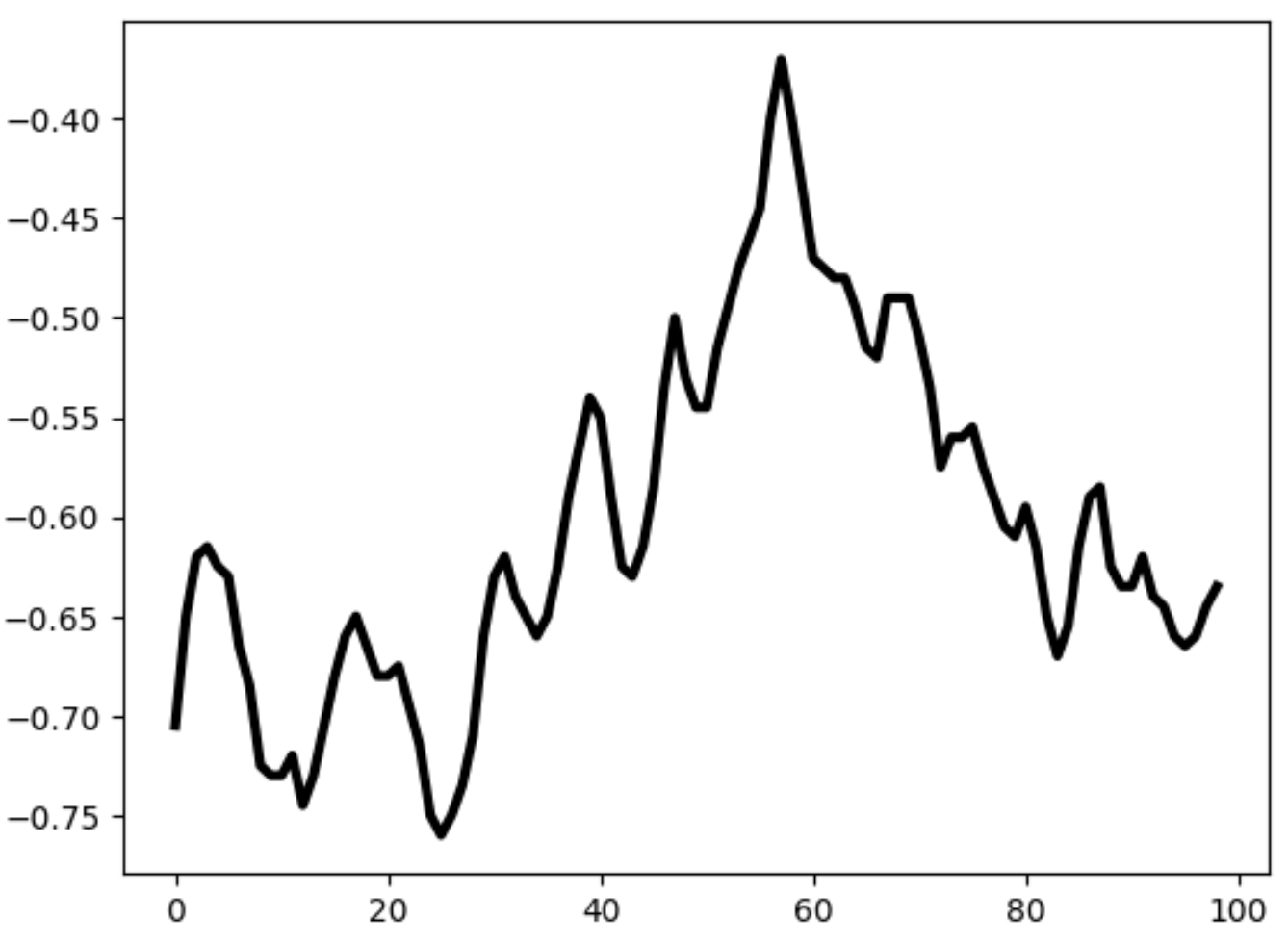}
    \caption{Lower outlier}
    \label{fig:out2}
  \end{subfigure}
  \caption{Example of outlier heartbeat}
  \label{fig:exampleOUT}
\end{figure}
Lower outliers are very short RR distances due to Analog-to-Digital Converter noise \cite{adcnoise} when recording the signals. Most of upper outliers are included in the normal class and represent only 1.8\% of total amount of data. 

The equations to calculate upper and lower outliers are as follows: 
\begin{equation}
Upper Outliers > Q_3 + 1.5 \times (Q_3-Q_1)\label{eq2}
\end{equation}
\begin{equation}
Lower Outliers < Q_1 - 1.5 \times (Q_3-Q_1)\label{eq3}
\end{equation}

$Q_1$ and $Q_3$ are respectfully the $25^{th}$ and $75^{th}$ percentile. Fig. \ref{fig:exampleOUT} shows a sample from both lower and upper outliers. About 5200 upper outliers and 1200 lower outliers were eliminated.

\subsection{Heartbeats and dataset creation} 
To extract heartbeats from the recordings, we first use a slope-sensitive QRS detector \cite{detector} to create a set of RR time intervals lengths from all the 48 (30-min) recordings. After that, we apply the following steps to generate the independent centered heartbeats:
\begin{enumerate}
    \item We retrieved the set of RR time intervals within each recording and eliminated any outlier intervals using the IQR method (see Section \ref{out}).
    \item We split each of the cleaned recordings (after outlier removal) into 10-seconds windows. 
    \item To generate new QRS-centered heartbeats, we calculated the optimal heartbeat size by averaging the RR intervals within current window. Then, we generated the new heartbeats centralized around their own R peaks.
\end{enumerate} 

\begin{figure}[htbp]
    \centering
    \includegraphics[width=7cm]{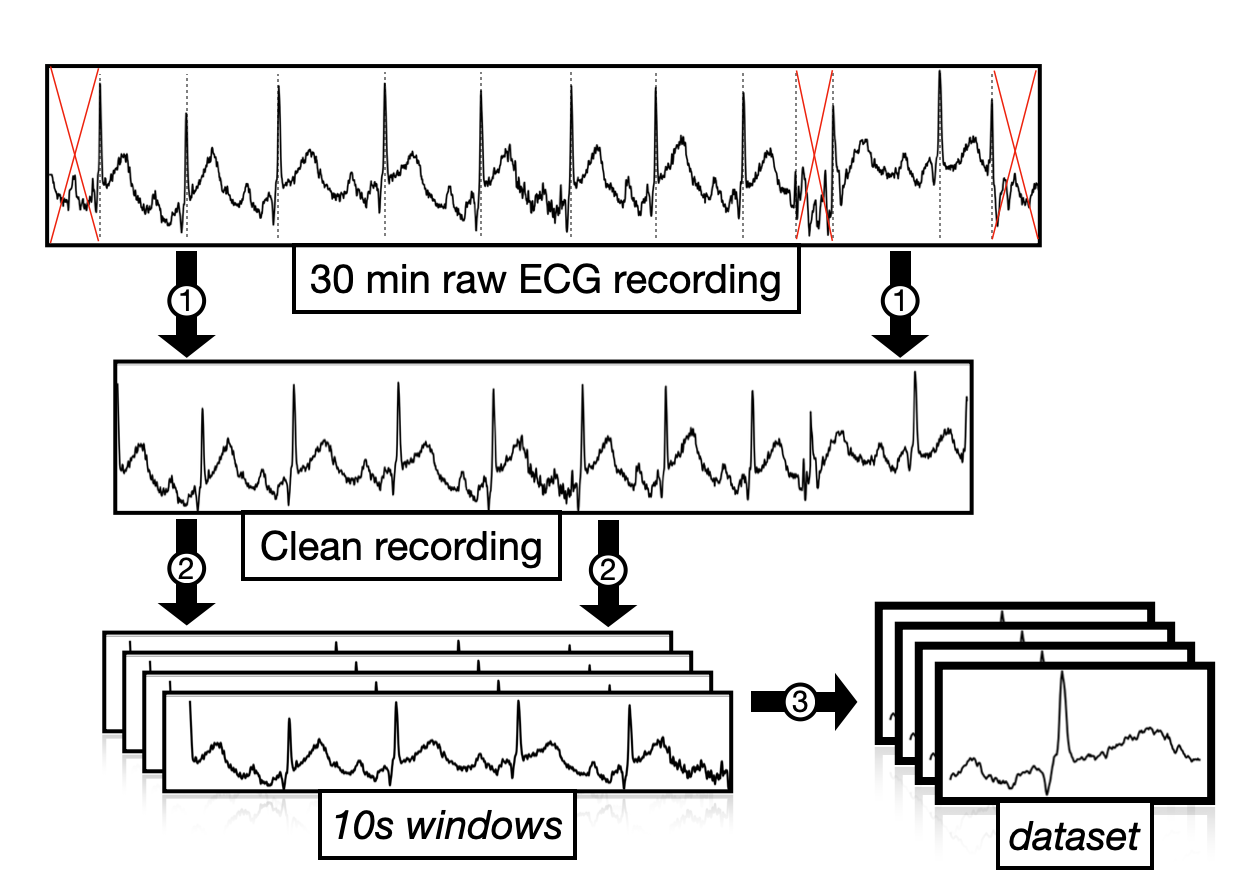}
    \caption{Dataset creation pipeline}
    \label{fig:datapipeline1}
\end{figure}

To create the dataset, it is necessary to ensure that all heartbeats have the same size. However, this process (Fig. \ref{fig:datapipeline1}) leads to a list of heartbeats with different sizes for each 10s window. To overcome this, we selected the maximum heartbeat size of 450 samples across all recordings as the \textit{Global heartbeat size} for the dataset. To match this \textit{global heartbeat size}, the remaining heartbeats have been padded with zeros resulting in 109,494 heartbeats to create the new dataset. This dataset is publicly available on GitHub page.

\subsection{Downsampling and Normalization}
\label{sec:downsample} 

To enhance the quality of the new dataset for heartbeat classification, we used downsampling and normalization. Heartbeats are mainly classified based on the signal's shape instead of bit-by-bit details. Hence, it is better to downsample heartbeats from 360 Hz to a sampling rate of 120 Hz. This results in 150 samples per heartbeat reducing memory usage by 3 times, as well as, impacting computational time, and complexity. Additionally, the dataset have been normalized using the z-score method \cite{normalize}.
%

%% file: sections/2model.tex
\section{Model architecture}

\subsection{Classification Model Architecture}
\begin{figure}[H]
    \centering
    \includegraphics[width=6.5cm]{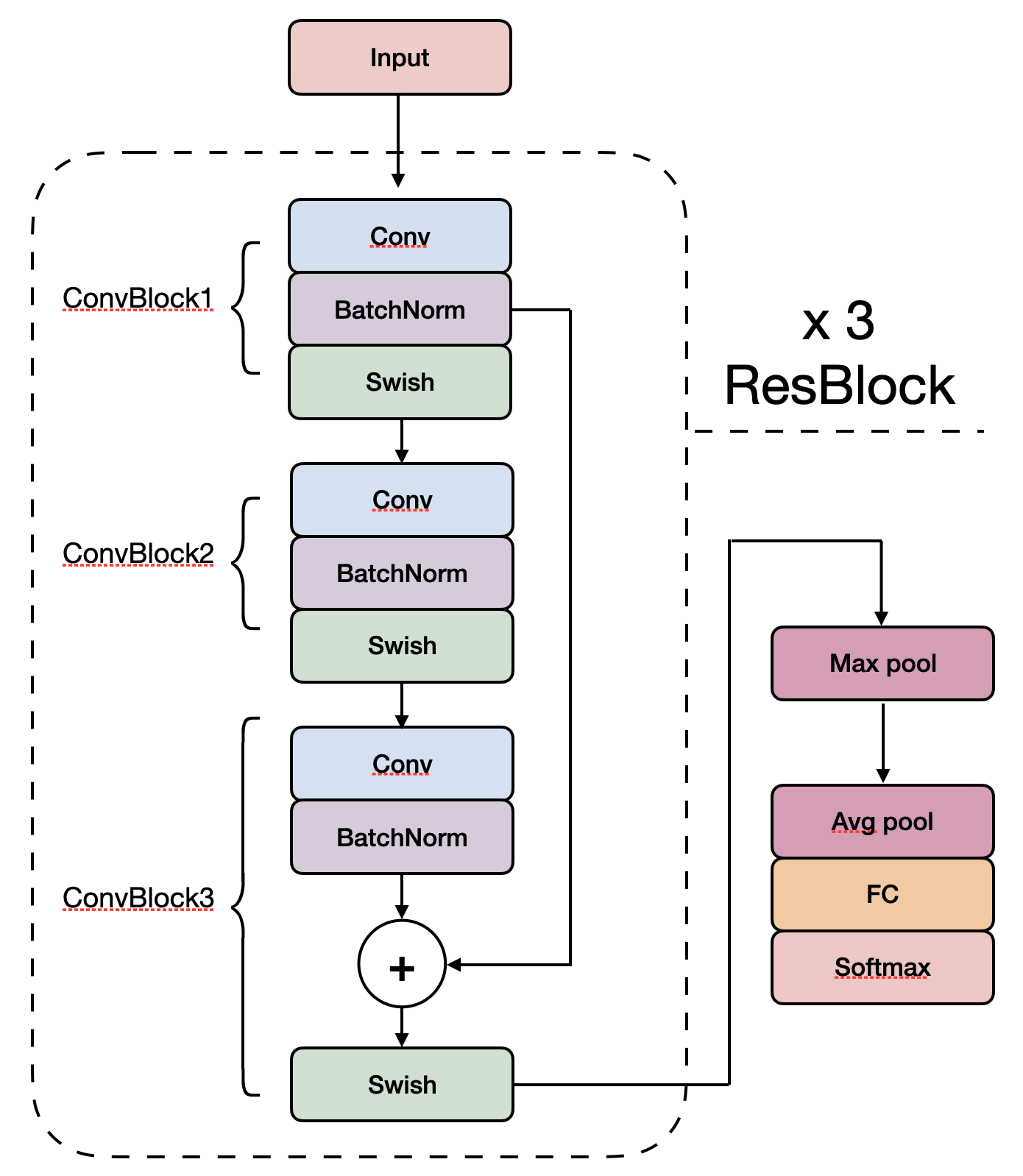}
    \caption{Model Architecture}
    \label{fig:model_arch}
\end{figure}

We train a 1-D residual convolutional neural network (ResNet) \cite{resnet} for ECG heartbeat classification, as shown in Fig. \ref{fig:model_arch}. The network comprises 34 layers, including 3 residual blocks. Each block contains 3 ConvBlocks, a skip connection, and a max pooling layer. Each ConvBlock includes a 1-D convolutional layer, a batch normalization layer, and a swish activation function. The number of convolution channels is 128, 64, and 32 for each residual block, respectively. After the last ResBlock, an AvgPooling layer computes an adaptive average over each channel, followed by a 32-neuron dense layer that outputs 5 classes normalized by a Softmax layer. This architecture, with only 269061 trainable parameters, is efficient and suitable for deployment on resource-constrained devices.

The proposed architecture is relatively deep and uses residual blocks, skip connections, and batch normalization to overcome vanishing gradients \cite{hochreiter1998vanishing} and improve gradient propagation, information flow, and training stability  \cite{resnet, batchnorm}. Additionally, the resulting model size of only 269061 trainable parameters suggests that the proposed architecture is efficient and may be suitable for deployment on resource-constrained devices such as real-time inference on an edge platform.

\subsection{Model Training}
We used the newly created dataset that consits of 109494 heartbeat signals, split into 80\% for training and 20\% for testing. The model was trained using categorical cross-entropy loss and the ADAM optimizer \cite{adam}, with a batch size of 512, accelerated by 2 Nvidia GTX 1070 GPUs. Training was performed using a custom batch generator with the Torch deep learning library \cite{torch}, with no data augmentation applied.

%% file: sections/3results.tex
\section{Results and discussion}

 Table \ref{tab:QRS_conf} shows 
the confusion matrix obtained by our training, the total number of samples in each class is indicated as support. 

\begin{minipage}{0.45\textwidth}
    \begin{table}[H] 
    \centering
        \caption{Model's confusion matrix on the test set of the new dataset}
    \hspace*{-0.9cm}\begin{tabular}{cccccccc}
    \multicolumn{2}{c}{} & \multicolumn{5}{c}{Predicted Class} &\\ \cline{2-8}
    \multicolumn{2}{c}{} & N & S & V & F & Q & Support\\ \cline{2-8}
    \multirow{5}{*}{\rotatebox[origin=c]{90}{Actual Class}} & N & \textbf{99.57} & 0.21 & 0.07 & 0.12 & 0 & 17920\\
    & S & 6.72 & \textbf{92.85} & 0.21 & 0.21 & 0 & 476\\
    & V & 1.36 & 0.14 & \textbf{97.62} & 0.86 & 0 & 1388\\
    & F & 7.54 & 0.62 & 1.88 & \textbf{89.93} & 0 & 159\\
    & Q & 0.53 & 0 & 0.17 & 0 & \textbf{99.29} & 1695\\ 
    
    \cline{2-8}
    \end{tabular}

    \label{tab:QRS_conf}
    \end{table}
\end{minipage}

\begin{table}[H]
    \centering
        \caption{Downsampling effect on model's performance}

    \begin{tabular}{cccc}
    \hline
    Preprocessing & Accuracy & Time (s/epoch) & Memory(MB/batch)\\
    \hline
    Raw & 98.68 & 123 & 1.76\\
    \textbf{Downsampling} & \textbf{99.24} & \textbf{82} & \textbf{0.59} \\ 
    \hline
    \end{tabular}  
    \label{tab:new}
\end{table}
Table \ref{tab:new} represents the impact of downsampling on the model's performance and training time. Since the ECG waves of the different classes are distinguishable by their general shapes as explained in \ref{sec:downsample}. This makes downsampling a promising preprocessing technique for this task. The results show that we achieved a $1.5\times$ speedup in training compared to the training of the raw data and a $2\%$ accuracy increase and reduced 3x memory usage.

\begin{table}[H]
    \centering
        \caption{Comparison of classification results with related papers}

    \begin{tabular}{c c c}
    \hline
    Work     &Approach      &Accuracy \\
    \hline
    \textbf{This work} &\textbf{1-D Resnet} &\textbf{99.2}\\
    Acharya et al \cite{ACHARYA2017389} &Augmentation + CNN &93.5\\
    Kachuee et al \cite{Kachuee} &Deep residual CNN &93.4\\

    \hline
    \end{tabular}  
    \label{tab:results200}
\end{table}

In Table \ref{tab:results200} the results of this paper are compared with \cite{ACHARYA2017389} and \cite{Kachuee}. 
We note that the accuracy achieved by this approach reached 99.2\%. however, accuracies published by  \cite{ACHARYA2017389}, \cite{Kachuee} are around 93\%. This means an improvement of 6\% due to the quality of the dataset and the deep architecture of the model.

Moreover, To ensure fair comparisons, we trained the  1-D Resnet model on the newly created dataset, as well as the dataset published by Kachuee et al. \cite{Kachuee}, and Acharya et al. \cite{ACHARYA2017389} (To use Acharya et al.'s dataset, we followed the methodology outlined in their paper since their dataset is not publicly available). We have selected the F class to evaluate the performance of each dataset, because it is the less populated class containing $0.72\%$ of the total data which makes it the most challenging and more likely to be misclassified by the model. Since the only variable in this comparison is the dataset, the results of table \ref{tab:F_class} reflect directly the performance of each dataset. 

\begin{table}[H]
    \centering
    \caption{Classification metrics for the F class with a $\pm3\%$ error, where the support represents the number of sample of the F class in the test set}
    \begin{tabular}{cccccc}
    \hline
    Dataset & Accuracy & Recall& Precision  & F-score& Support\\
    \hline
    \textbf{This work} & \textbf{87.26} &\textbf{87.25}&\textbf{96.2}& \textbf{90.95}&159\\
    Acharya et al. \cite{ACHARYA2017389} & 79.31 &79.31&90.54&84.54& 174\\
    Kachuee et al. \cite{Kachuee} & 72.83 & 72.81&95.87&82.86&162\\
    \hline
    \end{tabular}  
    
    \label{tab:F_class}
\end{table}

Results of Table \ref{tab:F_class} demonstrate that the use of our newly created dataset leads to improved performance, as our 1-D Resnet model achieved higher accuracy than when trained on the datasets from Kachuee \textit{et al.} and Acharya \textit{et al.} furthermore, our technique suggests that our dataset contains well-defined heartbeats and easily extractable features.

%% file: sections/4conc.tex
\section{Conclusion}

In this study, we presented a methodology for creating a high-quality heartbeat dataset based on the MIT-BIH recordings. We eliminated outlier heartbeats using the IQR method and created an optimal heartbeat size to overcome the problem of mixed heartbeats. The resulting dataset have been downsampled to accelerate training and reduce memory usage. We developed a 1-D Resnet architecture model to classify five different heartbeat types. The proposed method achieved 99.24\% accuracy with 5.7\% improvement compared to state-of-the-art methods \cite{ACHARYA2017389, Kachuee} due to the dataset. The use of outlier elimination and optimal window size selection highly contributed the quality of the dataset. The downsampling allowed for more efficient training accelerated the process by 33\% and 3x less memory usage. We will publish the dataset to facilitate further research in this area. In conclusion, this study highlights the effectiveness of our deep learning approach for ECG heartbeat classification and the importance of high-quality datasets for achieving accurate results.

%% file: main.bbl
\begin{thebibliography}{10}
\providecommand{\url}[1]{#1}
\csname url@samestyle\endcsname
\providecommand{\newblock}{\relax}
\providecommand{\bibinfo}[2]{#2}
\providecommand{\BIBentrySTDinterwordspacing}{\spaceskip=0pt\relax}
\providecommand{\BIBentryALTinterwordstretchfactor}{4}
\providecommand{\BIBentryALTinterwordspacing}{\spaceskip=\fontdimen2\font plus
\BIBentryALTinterwordstretchfactor\fontdimen3\font minus
  \fontdimen4\font\relax}
\providecommand{\BIBforeignlanguage}[2]{{%
\expandafter\ifx\csname l@#1\endcsname\relax
\typeout{** WARNING: IEEEtran.bst: No hyphenation pattern has been}%
\typeout{** loaded for the language `#1'. Using the pattern for}%
\typeout{** the default language instead.}%
\else
\language=\csname l@#1\endcsname
\fi
#2}}
\providecommand{\BIBdecl}{\relax}
\BIBdecl

\bibitem{who}
W.~H. Organization, \emph{World health statistics 2019: monitoring health for
  the SDGs, sustainable development goals}.\hskip 1em plus 0.5em minus
  0.4em\relax World Health Organization, 2019.

\bibitem{4015610}
O.~T. Inan, L.~Giovangrandi, and G.~T.~A. Kovacs, ``Robust neural-network-based
  classification of premature ventricular contractions using wavelet transform
  and timing interval features,'' \emph{IEEE Transactions on Biomedical
  Engineering}, vol.~53, no.~12, pp. 2507--2515, 2006.

\bibitem{5238547}
O.~Sayadi, M.~B. Shamsollahi, and G.~D. Clifford, ``Robust detection of
  premature ventricular contractions using a wave-based bayesian framework,''
  \emph{IEEE Transactions on Biomedical Engineering}, vol.~57, no.~2, pp.
  353--362, 2010.

\bibitem{7491263}
M.~Kachuee, M.~M. Kiani, H.~Mohammadzade, and M.~Shabany, ``Cuffless blood
  pressure estimation algorithms for continuous health-care monitoring,''
  \emph{IEEE Transactions on Biomedical Engineering}, vol.~64, no.~4, pp.
  859--869, 2017.

\bibitem{ACHARYA2017389}
\BIBentryALTinterwordspacing
U.~R. Acharya, S.~L. Oh, Y.~Hagiwara, J.~H. Tan, M.~Adam, A.~Gertych, and R.~S.
  Tan, ``A deep convolutional neural network model to classify heartbeats,''
  \emph{Computers in Biology and Medicine}, vol.~89, pp. 389--396, 2017.
  [Online]. Available:
  \url{https://www.sciencedirect.com/science/article/pii/S0010482517302810}
\BIBentrySTDinterwordspacing

\bibitem{Kachuee}
M.~Kachuee, S.~Fazeli, and M.~Sarrafzadeh, ``Ecg heartbeat classification: A
  deep transferable representation,'' in \emph{2018 IEEE International
  Conference on Healthcare Informatics (ICHI)}, 2018, pp. 443--444.

\bibitem{https://doi.org/10.48550/arxiv.1707.01836}
\BIBentryALTinterwordspacing
P.~Rajpurkar, A.~Y. Hannun, M.~Haghpanahi, C.~Bourn, and A.~Y. Ng,
  ``Cardiologist-level arrhythmia detection with convolutional neural
  networks,'' 2017. [Online]. Available: \url{https://arxiv.org/abs/1707.01836}
\BIBentrySTDinterwordspacing

\bibitem{mitbih}
G.~Moody and R.~Mark, ``The impact of the mit-bih arrhythmia database,''
  \emph{IEEE Engineering in Medicine and Biology Magazine}, vol.~20, no.~3, pp.
  45--50, 2001.

\bibitem{PTB}
A.~L. Goldberger, L.~A.~N. Amaral, L.~Glass, J.~M. Hausdorff, P.~C. Ivanov,
  R.~G. Mark, J.~E. Mietus, G.~B. Moody, C.-K. Peng, and H.~E. Stanley,
  ``Physiobank, physiotoolkit, and physionet: Components of a new research
  resource for complex physiologic signals.'' in \emph{Circulation}, 2000.

\bibitem{IQR}
A.~Turkyilmaz and B.~Alpar, ``Outlier detection methods: a review and
  comparative study,'' \emph{Expert Systems with Applications}, vol.~83, pp.
  157--173, 2017.

\bibitem{adcnoise}
R.~J. Halter, P.~B. Yoo, and M.~E. Josephson, ``Effect of analog-to-digital
  converter noise on electrocardiogram recordings,'' \emph{Journal of
  Cardiovascular Electrophysiology}, vol.~18, no.~4, pp. 402--410, 2007.

\bibitem{detector}
\BIBentryALTinterwordspacing
P.~S. Hamilton, ``Open source ecg analysis software documentation,'' QRS
  Detector, 2002. [Online]. Available:
  \url{https://physionet.org/content/physiotools/1.0.6/ecg-toolbox/qrs-detectors/slope/}
\BIBentrySTDinterwordspacing

\bibitem{normalize}
\BIBentryALTinterwordspacing
S.~G.~K. Patro and K.~K. Sahu, ``Normalization: {A} preprocessing stage,''
  \emph{CoRR}, vol. abs/1503.06462, 2015. [Online]. Available:
  \url{http://arxiv.org/abs/1503.06462}
\BIBentrySTDinterwordspacing

\bibitem{resnet}
K.~He, X.~Zhang, S.~Ren, and J.~Sun, ``Deep residual learning for image
  recognition,'' 06 2016, pp. 770--778.

\bibitem{hochreiter1998vanishing}
S.~Hochreiter, Y.~Bengio, P.~Frasconi, and J.~Schmidhuber, ``The vanishing
  gradient problem during learning recurrent neural nets and problem
  solutions,'' \emph{International Journal of Uncertainty, Fuzziness and
  Knowledge-Based Systems}, vol.~6, no.~02, pp. 107--116, 1998.

\bibitem{batchnorm}
S.~Ioffe and C.~Szegedy, ``Batch normalization: Accelerating deep network
  training by reducing internal covariate shift,'' \emph{International
  conference on machine learning}, pp. 448--456, 2015.

\bibitem{adam}
\BIBentryALTinterwordspacing
D.~P. Kingma and J.~Ba, ``Adam: A method for stochastic optimization,'' 2014.
  [Online]. Available: \url{https://arxiv.org/abs/1412.6980}
\BIBentrySTDinterwordspacing

\bibitem{torch}
\BIBentryALTinterwordspacing
A.~Paszke, S.~Gross, F.~Massa, A.~Lerer, J.~Bradbury, G.~Chanan, T.~Killeen,
  Z.~Lin, N.~Gimelshein, L.~Antiga, A.~Desmaison, A.~Köpf, E.~Yang, Z.~DeVito,
  M.~Raison, A.~Tejani, S.~Chilamkurthy, B.~Steiner, L.~Fang, J.~Bai, and
  S.~Chintala, ``Pytorch: An imperative style, high-performance deep learning
  library,'' 2019. [Online]. Available: \url{https://arxiv.org/abs/1912.01703}
\BIBentrySTDinterwordspacing

\end{thebibliography}
